# Effector Detection in Social Networks

Guangmo (Amo) Tong, *Student Member, IEEE*, Shasha Li, Weili Wu, *Member, IEEE*, and Ding-Zhu Du, *Member, IEEE*

*Abstract*—In a social network, influence diffusion is the process of spreading innovations from user to user. An activation state identifies who are the active users who have adopted the target innovation. Given an activation state of a certain diffusion, effector detection aims to reveal the active users who are able to best explain the observed state. In this paper, we tackle the effector detection problem from two perspectives. The first approach is based on the influence distance that measures the chance that an active user can activate its neighbors. For a certain pair of users, the shorter the influence distance, the higher probability that one can activate the other. Given an activation state, the effectors are expected to have short influence distance to active users while long to inactive users. By this idea, we propose the influence-distance-based effector detection problem and provide a 3-approximation. Second, we address the effector detection problem by the maximum likelihood estimation (MLE) approach. We prove that the optimal MLE can be obtained in polynomial time for connected directed acyclic graphs. For general graphs, we first extract a directed acyclic subgraph that can well preserve the information in the original graph and then apply the MLE approach to the extracted subgraph to obtain the effectors. The effectiveness of our algorithms is experimentally verified via simulations on the real-world social network.

## I. INTRODUCTION

**S**OCIAL networks serve as important platforms for information propagation as they allow efficient and effective communications between participants. In the sense of online social networks, e.g., Flicker, Facebook, and Twitter, new topics may spread rapidly through the networks via the friendship relation. In the context of viral marketing, companies employ P2P advertising to achieve the increase in brand awareness and sales. At the same time, risks from rumors or viruses also propagate through the Internet or local area network. During the last two decades, the problems regarding influence diffusion has been extensively studied in many domains, such as economics, epidemiology, and social media.

For a social network with a certain influence diffusion, an activation state shows the users who have been influenced. The effectors are the nodes that can best explain the observed activation state. That is, *once selected as seed nodes, the*

Manuscript received May 29, 2016; revised September 12, 2016; accepted November 6, 2016. This work was supported in part by the National Science Foundation of China (No. 11301480) and the Natural Science Foundation of Ningbo, China (No. 2014A610030).

G. Tong, W. Wu, and D.-Z. Du are with the Department of Computer Science, Erik Jonsson School of Engineering and Computer Science, University of Texas at Dallas, Richardson, TX 75080 USA (e-mail: gxt140030@utdallas.edu).

S. Li is with the Ningbo Institute of Technology, Zhejiang University, Ningbo 315100, China.

Digital Object Identifier 10.1109/TCSS.2016.2627811

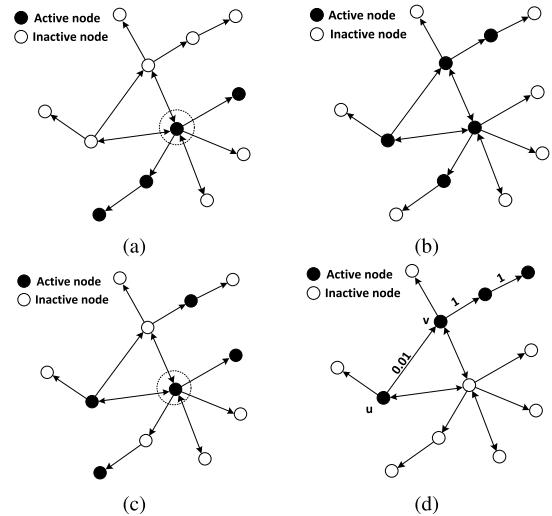

Fig. 1. Example network. (a) Example. (b) Example. (c) Example. (d) Example.

*effectors are able to activate exactly the active nodes specified by the given activation state*. It has been shown that the identification of effectors is of great significance in understanding the dynamics of influence diffusion. On one hand, effectors play as the critical nodes that encourage or halt the spread of new ideas and technologies. On the other hand, effectors may be interpreted as the source of the influence propagation, and thus effector detection helps in identifying the users who are the inventors of novel ideas or the culprits initiating rumors or infections. Nevertheless, measuring the quality of effector is not a trivial problem. Suppose there is one effector. For the activation state shown in Fig. 1(a), the circled node can be a good effector since it is the only node that is able to activate all other active nodes. However, for the second activation state shown in Fig. 1(b), determining the best effector from the active nodes requires more investigation. In fact, the criteria for selecting effectors depend on not only the pattern of the activation state but also on the diffusion model. We will formally define our problem in Section III.

The problem of identifying propagation sources in networks has been widely investigated by Jiang *et al.* [1], while effector detection has received much less attention. It is worthy to note that these two problems are, although similar, not identical. The propagation source identification problem assumes that the given activation state is a spreading result caused by a certain seed set, while the activation state in the effector detection problem does not necessarily have this property. For example, assuming that the budget is one, the activation state shown in Fig. 1(c) is not a valid input for the propagation source identification problem since there is no node that can result such an influence pattern. However, we can still solve the effector detection problem on this activation state as there always exists the node that can produce an activation state that is the most similar to the given one. Furthermore, the solutions to these problems may not be the same for the same input. Consider the example shown in Fig. 1(d). Suppose associated with each directed edge $(u, v)$ there is a weight $w(u, v) \in [0, 1]$, which denotes the probability that $u$ can activate $v$ after becoming active. For this activation state, the source of the influence should be node $u$, but node $v$ is actually a better choice for selecting effector. One can check that for the four active nodes in Fig. 1(d), $u$ activates 1.02 of them on average, while $v$ brings three active nodes for sure. Therefore, techniques designed for identifying influence sources may not be applicable for detecting good effectors. Furthermore, most of the prior works are focused on epidemic models such as susceptible–infected (SI) model, SI–susceptible (SIS) model, and SI–recovery (SIR) model, and a few works have considered the information–propagation models for social networks. In this paper, we consider a classic information–propagation model, namely, independent cascade (IC) model.

Lappas *et al.* [2] are among the first who studied the effector detection problem for an IC model. In particular, they propose the effector problem from the perspective of combinatorial optimization and prove that their problem is NP-hard to be solved optimally or efficiently approximated. As expected, the formulation of the problem determines the computational difficulty. In this paper, following their work, we consider the IC model but formulate the effector detection problem from different views.

### A. Our Contribution

In a social network, influence distance is a metric that measures the distance between users with respect to information propagation. That is, information is more easily to spread from one user to another if the influence distance is short. Therefore, given an activation state, the influence distance between the effectors and active (also inactive) nodes should be short (also long). As shown in the experiments, given an activation state, for a user set $S$, there is a clear negative (also positive) correlation between the effector quality of $S$ and the influence distance between $S$ and active (also inactive) nodes. According to that, we introduce the influence-distance-based effector detection problem. For this problem, we first show a 3-approximation and then provide an efficient heuristic. Our second method for selecting effectors employs the approach of maximum likelihood estimation (MLE). One can see that the nodes that, if selected as seed nodes, are able to maximize the probability of observing the given activation state are naturally to be good effectors. For the directed acyclic graphs (DAGs), we show that the optimal MLE can be obtained in polynomial. For general graphs, we proceed in two steps. We first extract a DAG subgraph that maximally preserves the information in the original graph with respect to entropy and then apply the MLE approach to the extracted DAG to detect effectors. By the results of the experiments conducted on the real social network, both of our approaches can produce high-quality effectors and perform better than the state-of-art method.

The rest of this paper is organized as follows. Section II reviews the related work. In Section III, we introduce the IC model and formally define the metric of effector quality. The proposed two approaches are provided in Sections IV and V, respectively. In Section VI, we experimentally evaluate our methods using the Facebook network. Section VII concludes this paper.

## II. RELATED WORK

Lappas *et al.* [2] proposed the effector detection problem for the IC model. According to their formulation, the optimal solution is obtainable on tree graphs in polynomial time, while the general graphs are hard to handle. They also propose several heuristics and explore their performance via experiments. We are not aware of any other work addressing this problem for social networks, but there has been a huge body of works regarding the identification of influential nodes or propagation sources.

The topics regarding selecting important users in social networks have been considered for different application domains. Domingos and Richardson [9] first studied the influential customers in viral marketing. Motivated by [9], Kempe *et al.* [10] propose the well-known influence maximization problem that aims to select the seed users that can maximize the influence. Following their framework, many relevant problems in social network have been studied [11]–[14]. In the influence maximization problem, the goal is to influence the nodes as many as possible while the effector detection problem focuses on achieving a certain activation state. It has been shown that the influence maximization problem can be taken as a special case of the effector detection problem. In the context of immunization, the important nodes are the best for removal to halt the spread of epidemics [15]–[17]. One can see that the immunization problem is the opposite problem to the influence maximization problem, and, in fact, the effector detection problem is somehow the combination of these two problems. We will later discuss this in Section III. Rusmevichientong *et al.* study the problem of finding the consumers who tend to purchase products earlier than others. According to their formulation, the early buyers correspond to the nodes with large difference between the weights of the outgoing and incoming edges. Different from our setting, their network is constructed based on the purchase information and they do not consider the influence dynamics.

A series of methods for identifying propagation sources has been proposed for epidemic models, e.g., SI, SIR, and SIS models. In such models, the nodes are initially susceptible and may later switch states among susceptible, infected, or recovery. In the SI model, a node never changes its state once becoming infected, while in the SIR or SIS mode, the infected node can later become recovery or susceptible again. Shah and Zaman [3] study the problem of detecting sources of computer viruses in the SI model based on the concept of rumor centrality. Dong *et al.* [4] consider the single source detection problem by constructing the maximum *a posteriori* estimator. The similar methods have also been applied to SIS and SIR models [5], [6]. Most of the initial works focus on the tree-like networks to obtain theoretical results. Techniques for general graphs appear in recent years. Prakash *et al.* [7] consider the generic network topology and propose the NETSLEUTH method for the SI model. Later, Altarelli *et al.* [8] study several Bayesian inference problems for irreversible stochastic epidemic models, namely, the SIR model. In another issue, the propagation sources can be detected by injecting sensors. When the sensors are injected into networks, they act as normal users while collecting the propagation information including states, transition time, and infection directions. As mentioned in Section I, the propagation source detection problem is not identical to the effector detection problem. The effectors are the nodes that can result in the activation state similar to the one given, while the sources are the nodes initiating the influence diffusion.

## III. PROBLEM

In this section, we give the preliminaries regarding the IC model and discuss the issue on defining good effectors. In this section, we give the preliminaries regarding the IC model and discuss the issue on defining good effectors. The notations that are frequently used in the rest of this paper are listed in Table I.

A social network is represented by a directed graph $G = (V, E)$. Let $V = \{u_1, \ldots, u_N\}$, where $N$ is the number of users in $G$. In order to initiate the influence diffusion, we first activate a set of seed users who will potentially activate their friends. We speak of each user as being either *active* or *inactive*. In the IC model, associated with each edge $e = (u, v)$, there is a $\Pr[e] \in [0, 1]$, which is the *propagation probability* from $u$ to $v$. An active node $u$ has only one chance to activate its inactive neighbor $v$ with the probability of $\Pr[(u, v)]$. If $u$ successfully activates $v$ via edge $(u, v)$, we call edge $(u, v)$ a *live* edge. After the activation of seed users, the process of influence diffusion goes round by round and terminates when there is no user can be further activated. An *activation state* $\mathbf{A} = (a_1, \ldots, a_N)$ is a vector of $\{0, 1\}^N$, where $a_i = 1$ (also $a_i = 0$) if user $i$ is active (also inactive). Given an activation state $\mathbf{A}$, we denote the set of active (also inactive) nodes in $A$ by $X_1^A$ (also $X_0^A$). For a node set $S$ that is selected as seed set, we denote by $\mathbf{A}_S$ the activation state when the influence diffusion terminates. Note that $\mathbf{A}_S$ is a random vector since the IC model is a probabilistic model. Given an integer $B$, the $B$-effector detection problem is defined as follows.

Given a target activation state $\mathbf{A}^*$, we aim to find a subset[1] $S$ of $X_1^{\mathbf{A}^*}$ with $B$ nodes, which is able to result an activation state $\mathbf{A}_S$ similar to $\mathbf{A}^*$.

Since the given state $\mathbf{A}^*$ is fixed, in the rest of this paper, we will refer to $X_1^{\mathbf{A}^*}$ (also $X_0^{\mathbf{A}^*}$) as $X_1$ (also $X_0$). Let $N_1 = |X_1|$ be the number of nodes in $X_1$. Let $\|\cdot\|$ be the $L^1$-norm and $E[\cdot]$ be the expected value of a random variable. In this paper, we consider the following metrics to measure the quality of a node set $S$. The first measure is given by

$$f_1(S) = E[\|\mathbf{A}^* - \mathbf{A}_S\|] \qquad (1)$$

[1]Since the effectors are naturally active, they are limited in $X_1^{\mathbf{A}^*}$.

### TABLE I
### NOTATIONS

| Symbol | Definition |
|---|---|
| $G = (V, E)$ | instance of IC network. |
| $N$ | $N = |V|$. |
| $\Pr[e]$ | propagation probability of edge $e$. |
| $\overline{P}_{uv}^G$ | the maximum diffusion path from $u$ to $v$. |
| $P_{uv}^k$ | the $k$-max path set from $u$ to $v$. |
| $d_{(u,v)}^k$ | the k-th influence distance. |
| $\mathbf{A}^*$ | target activation state. |
| $X_1$ | set of the active nodes in $\mathbf{A}^*$. |
| $X_0$ | set of the inactive nodes in $\mathbf{A}^*$. |
| $N_1$ | $N_1 = |X_1|$. |
| $B$ | number of effectors. |
| $f_1(S), f_2(S)$ | two metrics of effector quality . |
| $g_k(S)$ | the objective function of the k-IDBED problem. |
| $G_{uv}$ | the bipartite graph constructed by Algorithm 1. |
| $\Pr(\mathbf{A}|S)$ | the probability of observing $\mathbf{A}$ given a seed set $S$. |

which is the expected difference between the target activation state $\mathbf{A}^*$ and the one resulted by $S$. Let $\alpha(S, u)$ be the probability that node $u$ can be activated when $S$ is selected as the seed set. The second measure is defined as follows:

$$f_2(S) = \|\mathbf{A}^* - E[\mathbf{A}_S]\| \qquad (2)$$

where $E[\mathbf{A}_S] = (\alpha(S, u_1), \ldots, \alpha(S, u_n))$ is the expected activation state resulted by $S$. Note that when $\mathbf{A}^* = (1, \ldots, 1)$, our problem reduces to the influence maximization problem, which aims to find a set of seed users to maximize the number of final active users. As shown in [2], it is NP-hard to minimize (1) or (2) and we cannot even obtain an $\alpha$-approximation for $\alpha > 1$ unless NP $= P$. In this paper, we present two techniques to find a node set $S$, $S \subseteq X_1$, with small $f_1(S)$ and $f_2(S)$.

## IV. INFLUENCE-DISTANCE-BASED EFFECTOR DETECTION

Our first approach is designed based on the concept of influence distance. Generally speaking, for two users $u$ and $v$, the influence distance $d(u, v)$ measures the chance that the influence can be spread to $v$ from $u$. The larger the value of $d(u, v)$, the lower the probability with which $u$ can activate $v$. Therefore, given the activation state $\mathbf{A}^*$, the nodes in $X_1$ (also $X_0$) are expected to have short (also long) distance to the effectors. From this perspective, we formulate the effector detection problem as an optimization problem and provide a 3-approximation algorithm.

We start by discussing how to extract the influence distance from the IC model. A path $pp_{uv}$ from node $u$ to $v$ is a set of edges and its propagation probability $\Pr[pp_{uv}]$ is defined as $\prod_{e \in pp_{uv}} \Pr[e]$. Intuitively, $u$ can activate $v$ via path $pp_{uv}$ with the probability of $\Pr[pp_{uv}]$. Let $\overline{pp}_{uv}^G$ be the path in $G$ that has the maximum path propagation probability from $u$ to $v$ and we call $\overline{pp}_{uv}^G$ the *maximum diffusion path* from $u$ to $v$. Clearly, $\overline{pp}_{uv}^G$ cannot completely capture the influence diffusion from $u$ to $v$ as there are many alternative joint paths. However, if we take the correlation of all the paths into consideration and calculate $\alpha(\{u\}, v)$ precisely, we will encounter a new problem that is #P-hard [18]. We herein develop the $k$th influence distance $d_{(u,v)}^k$ that considers the $k$ independent paths from $u$ to $v$ that have the largest propagation probability. For two users $u$ and $v$, the $k$th influence distance is induced by the $k$-max path set $P_{uv}^k$ defined as follows:

$$P_{uv}^k = \begin{cases} P_{uv}^{k-1} \cup \{\overline{pp}_{uv}^{G_k}\} & \text{if } k > 1 \\ \{\overline{pp}_{uv}^{G_1}\} & \text{if } k = 1 \end{cases}$$

and

$$G_k = \begin{cases} (V(G_{k-1}), E(G_{k-1}) \setminus \overline{pp}_{uv}^{G_{k-1}}) & \text{if } k > 1 \\ G & \text{if } k = 1 \end{cases}$$

where $G_k$ is a series of instances of IC model and $V(G)$ and $E(G)$, respectively, denote the vertex set and edge set of $G$. Recall that $\overline{pp}_{uv}^{G_{k-1}}$ is the maximum diffusion path from $u$ to $v$ in $G_k$. In the above process, we successively find the maximum diffusion path in $G_k$ and $G_{k+1}$ is obtained from $G_k$ by removing $\overline{pp}_{uv}^{G_k}$. The $k$th influence distance $d_{(u,v)}^k$ of the ordered pair of nodes $u$ and $v$ is computed by

$$d_{(u,v)}^k = -\ln\left(1 - \prod_{pp_{uv} \in P_{uv}^k} (1 - \Pr[pp_{uv}])\right). \tag{3}$$

By the construction of $P_{uv}^k$, the paths in $P_{uv}^k$ are edge disjoint and thus $u$ activates $v$ independently through these paths. Therefore, $1 - \prod_{pp_{uv} \in P_{uv}^k} (1 - \Pr[pp_{uv}])$ represents the probability that $u$ can activate $v$ through the paths in $P_{uv}^k$. Note that $1 - \prod_{pp_{uv} \in P_{uv}^k} (1 - \Pr[pp_{uv}])$ is a lower bound of $\alpha(\{u\}, v)$ and it is more closer to $\alpha(\{u\}, v)$ when $k$ is large. One can also see that a small $d_{(u,v)}^k$ implies the high propagation probability between $u$ and $v$. For a node set $V'$ and a node $u$, the $k$th influence distance $d_{(V',u)}^k$ between $V'$ and $u$ is defined as $\sum_{v \in V'} d_{(v,u)}^k$ and for two sets $V_1$ and $V_2$, $d_{(V_1, V_2)}^k = \sum_{u \in V_1} \sum_{v \in V_2} d_{(u,v)}^k$. Note that $d_{(u,v)}^k$ may not equal to $d_{(v,u)}^k$ as we consider the directed graph. Now let us go back to the effector detection problem. Given an activation state $\mathbf{A}^*$, we expect a set $S$ of effectors such that the influence distance between $S$ and $u$ is small (also large) if $u$ is active (also inactive) in $\mathbf{A}^*$. Thus, the solution to the following problem serves as a good effector set:

$$\min \; \lambda \cdot \sum_{u \in S} \sum_{v \in X_1 \setminus S} d_{(u,v)}^k - (1 - \lambda) \cdot \sum_{u \in S} \sum_{v \in X_0} d_{(u,v)}^k$$
$$\text{s.t. } |S| = B$$
$$S \subseteq X_1$$

where $\lambda \in [0, 1]$ is an adjustable parameter. The terms $\sum_{u \in S} \sum_{v \in X_1 \setminus S} d_{(u,v)}^k$ and $\sum_{u \in S} \sum_{v \in X_0} d_{(u,v)}^k$ are the distance from $S$ to active and inactive nodes, respectively. If $\lambda$ is close to 1, it implies that making the nodes in $X_1$ active is a prime consideration and it is not important whether the nodes in $X_0$ are inactive.

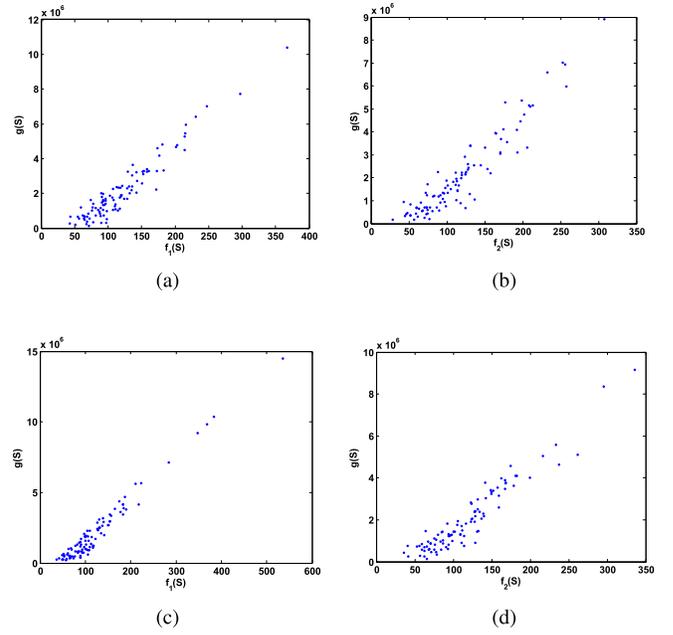

Fig. 2. Verifying the effectiveness of $g_k(S)$. In all four graphs, the $y$-axis and $x$-axis denote the objective value and the score of the quality metric of the generated effectors, respectively. Each graph gives four curves plotting the influence spread under four seeding strategies, respectively. (a) $f_1(S)$ versus $g_1(S)$. (b) $f_2(S)$ versus $g_1(S)$. (c) $f_1(S)$ versus $g_3(S)$. (d) $f_2(S)$ versus $g_3(S)$.

Given $X_0$, $X_1$, and $S$, the whole network is partitioned into three parts, $X_0$, $X_1 \setminus S$, and $S$. Since

$$\sum_{u \in S} \sum_{v \in X_0} d_{(u,v)}^k = d_{(X_1, X_0)}^k - d_{(X_1/S, X_0)}^k \tag{4}$$

and $d_{(X_1, X_0)}^k$ is fixed, the above problem is identical to the following one.

*Problem 1 (k-IDBED):*

$$\min \; g_k(S) = \lambda \cdot d_{(S, X_1/S)}^k + (1 - \lambda) \cdot d_{(X_1/S, X_0)}^k$$
$$\text{s.t. } |S| = B$$
$$S \subseteq X_1.$$

We denote the above problem by the $k$th influence-distance-based effector detection (k-IDBED) problem. Before discussing how to solve this problem, we first show the relationship between the objective value $g_k(S)$ and the quality of effectors according to $f_1(S)$ and $f_2(S)$. The simulation results evince that for a node set $S$, a small $g_k(S)$ implies a high quality of $S$. We employ the Facebook network [19], which contains 4036 nodes and 176 468 edges. The influence propagation probability for an edge $(u, v)$ is set as $1/\deg(u)$, where $\deg(u)$ is the out-degree of node $u$. Here we set $\lambda$ to be 0.5. We generate 1000 activation states where each activation state is generated by randomly selecting ten seed nodes and simulating the diffusion process. We set $k$ to be 1 and 3, respectively. For each activation state, we randomly generate an effector set $S$ with ten nodes and calculate $g_k(S)$, $f_1(S)$ and $f_2(S)$. The scatter diagrams are plotted in Fig. 2. Since small $f_1(S)$ or $f_2(S)$ implies high effector quality, the results strongly support that the node set $S$ with small $g_k(S)$ can be good effectors.

In the following, we study the cases for $k = 1$ and $k > 1$, respectively.

### A. 1-IDBSD Problem

In this section, we present a 3-approximation algorithm for Problem 1 when $k = 1$. In particular, the 1-IDBSD problem is stated as follows.

*Problem 2 (1-IDBED Problem):*

$$\min g_1(S) = \lambda \cdot d^1_{(S, X_1/S)} + (1 - \lambda) \cdot d^1_{(X_0, X_1/S)}$$
$$\text{s.t.} \ |S| = B$$
$$S \subseteq X_1.$$

Before presenting the algorithm, we introduce some necessary preliminaries.

*1) Triangle Inequality:* When $k = 1$, $d^k_{(u,v)}$ has the following property.

*Lemma 1:* For every three nodes $u$, $v$, and $w$ in $V$

$$d^1_{(u,v)} + d^1_{(v,w)} \geq d^1_{(u,w)}. \tag{5}$$

*Proof:* By (3)

$$d^1_{(u,v)} = -\ln\left(1 - \prod_{pp_{uv} \in P^1_{uv}}(1 - \Pr[pp_{uv}])\right)$$
$$= -\ln\left(\Pr[\overline{pp}^G_{uv}]\right). \tag{6}$$

Thus

$$d^1_{(u,v)} + d^1_{(v,w)} = -\ln\left(\Pr[\overline{pp}^G_{uv}]\right) - \ln\left(\Pr[\overline{pp}^G_{vw}]\right)$$
$$= -\ln\left(\Pr[\overline{pp}^G_{uv}] \cdot \Pr[\overline{pp}^G_{vw}]\right).$$

Note that $\overline{pp}^G_{uv}$ combining with $\overline{pp}^G_{vw}$ is a path from $u$ to $w$ with the propagation probability of $\Pr[\overline{pp}^G_{uv}] \cdot \Pr[\overline{pp}^G_{vw}]$. Thus, by the selection of $\overline{pp}^G_{uw}$, $\Pr[\overline{pp}^G_{uv}] \cdot \Pr[\overline{pp}^G_{vw}]$ is less than $\Pr[\overline{pp}^G_{uw}]$, and therefore

$$d^1_{(u,v)} + d^1_{(v,w)} \geq -\ln\left(\Pr[\overline{pp}^G_{uw}]\right) = d^1_{u,w}.$$

$\square$

The above triangle inequality allows us to design efficient approximation algorithms for Problem 2. As shown later in this section, the perfect matching problem plays an important role in solving Problem 2.

*2) Bipartite Minimum Perfect Matching Problem:* Given a weighted complete bipartite graph $G = (V_1, V_2, E)$ with a weight $w(e)$ on each edge $e \in E$, the problem asks for a perfect matching with the minimum weight. It is well known that this problem can be solved in polynomial time via linear programming. We assume there is a solver denoted by BMPM(G) whenever we need to solve this problem.

For an ordered pair $(u, v)$ of two nodes $u$ and $v$ in $X_1$, we construct a complete bipartite graph as follows.

*3) Bipartite Graph Construction:* Given two nodes $u, v \in X_1$, we construct a weighted complete bipartite graph $G_{uv} = (V_1, V_2, E)$ with a weight function $w_{uv}(\cdot)$ as follows. $V_1$ and $V_2$ both have $N_1$ nodes, where $V_1 = X_1$

---

**Algorithm 1** Bipartite Construction $(G, X_1, u, v, N_1, B)$

1: **Input:** $G = (V, E)$, $X_1$, $u_i \in X_i$, $u_j \in X_j$, $B$ and $\lambda$.
2: **Output:** A complete bipartite graph $(V_1, V_2, E)$ with a weight function $w(\cdot)$.
3: $N_1 := |X_1|$;
4: $X_0 := V \setminus X_1$;
5: $V_1 := X_1$;
6: $V_2 := \{v_2^1, \ldots, v_2^{N_1}\}$;
7: **for** each $w \in V_1$ **do**
8:     **for** $i = 1 : N_1$ **do**
9:         **if** $i \leq B$ **then**
10:             Set $w(w, v_2^i) = \lambda \cdot (N_1 - B) \cdot d^1_{(w,v)}$;
11:         **else**
12:             Set $w(w, v_2^i) = \lambda \cdot B \cdot d^1_{(v,w)} + (1 - \lambda) \cdot d^1_{(w, X_0)}$;
        **return** $(V_1, V_2, E)$ with $w(\cdot)$.

---

and $V_2 = \{v_2^1, \ldots, v_2^{N_1}\}$. For each $v' \in V_1$ (i.e., $X_1$), the weight function is defined as follows:

$$w_{uv}(v', v_2^i)$$
$$= \begin{cases} \lambda \cdot (N_1 - B) \cdot d^1_{(v', u)} & \text{if } 1 \leq i \leq B \\ \lambda \cdot B \cdot d^1_{(v', v)} + (1 - \lambda) d^1_{(v', X_0)} & \text{if } B < i \leq N_1. \end{cases}$$

The above construction is formally shown in Algorithm 1. For a perfect matching of $G_{uv}$, each node in $V_1$ is matched to one node in $V_2$, and thus there are exactly $B$ nodes matched to the nodes in $\{v_2^1, \ldots, v_2^B\} \subseteq V_2$. For each pair of nodes $u$ and $v$ in $G$, let $E_{uv}$ be the edge set of the minimum perfect matching of $G_{uv}$, $V'_{uv}$ be the set of the nodes matched to $\{v_2^1, \ldots, v_2^B\} \subseteq V_2$ in the minimum perfect matching, and $W(E_{uv})$ be the total weight of the edges in $E_{uv}$. By the construction of $G_{uv}$

$$W(E_{uv}) = \sum_{v^* \in V'_{uv}} \lambda \cdot (N_1 - B) \cdot d^1_{(v^*, u)}$$
$$+ \sum_{v^* \in X_1 \setminus V'_{uv}} \left(\lambda \cdot B \cdot d^1_{(v,v^*)} + (1 - \lambda) \cdot d^1_{(v^*, X_0)}\right).$$
$$= \lambda \cdot (N_1 - B) \cdot d^1_{(V'_{uv}, u)} + \lambda \cdot B \cdot d^1_{(v, X_1 \setminus V'_{uv})}$$
$$+ (1 - \lambda) \cdot d^1_{(X_1 \setminus V'_{uv}, X_0)}.$$

Now we are ready to solve Problem 2. Consider the algorithm shown in Algorithm 2, named the matching-based effector detection (MBED) algorithm. In this algorithm, for each pair of nodes $u, v \in X_1$, we first construct a bipartite graph $G_{uv}$ according to Algorithm 1 and then calculate its minimum perfecting matching. In this way, associated with each pair of nodes $u$ and $v$ in $X_1$, there is a node set $V'_{uv}$ and an edge set $E_{uv}$ given by the minimum perfect matching of $G_{uv}$. Finally, we select the $V'_{uv}$ such that $W(E_{uv}) + B \cdot (N_1 - B) \cdot d^1_{(u,v)}$ is minimized. Suppose the output of MBED is $V'_{u^*v^*}$ corresponding to a certain pair of nodes $u^*$ and $v^*$ in $X_1$. In the following, we show that $V'_{u^*v^*}$ is a 3-approximation of Problem 2.

*Lemma 2:* For any $V' \subseteq X_1$ with $B$ nodes and any pair of nodes $u, v \in X_1$

$$h\left(V'_{u^*v^*}, u^*, v^*\right) \leq h(V', u, v) \tag{7}$$

---

**Algorithm 2** Matching-Based Effector Detection

1: **Input**: $G = (V, E)$, $X_1$, $N_1$, $B$ and $\lambda$.
2: **Output**: A subset of $X_1$ with $B$ nodes.
3: $N_1 := |X_1|$.
4: **for** each ordered pair of $(u, v) \in X_1, u \neq v$, **do**
5:     $G_{uv} :=$ Bipartite-construction $(G, X_1, u, v, B, \lambda)$;
6:     run BMPM($G_{uv}$) to obtain $E_{uv}$ and $V'_{uv}$;
7: Return the $V'_{uv}$ with the smallest $W(E_{uv}) + \lambda \cdot B \cdot (N_1 - B) \cdot d^1_{(u,v)}$.

---

where

$$h(V', u, v) = \lambda \cdot (N_1 - B) \cdot d^1_{(V',u)} + \lambda \cdot B \cdot d^1_{(v, X_1 \setminus V')}$$
$$+ (1 - \lambda) \cdot d^1_{(X_1 \setminus V', X_0)} + \lambda \cdot B \cdot (N_1 - B) \cdot d^1_{(u,v)} \quad (8)$$

is a function defined on the triple $(V', u, v)$.

*Proof:* According to (8) and the construction of $G_{u^*v^*}$

$$h(V'_{u^*v^*}, u^*, v^*) = W(E_{u^*v^*}) + \lambda \cdot B \cdot (N_1 - B) \cdot d^1_{(u^*,v^*)}.$$

Consider the bipartite graph $G_{u,v}$ constructed according to Algorithm 1. By the selection of $u^*$ and $v^*$

$$W(E_{u^*v^*}) + \lambda \cdot B \cdot (N_1 - B) \cdot d^1_{(u^*,v^*)}$$
$$\leq W(E_{uv}) + \lambda \cdot B \cdot (N_1 - B) \cdot d^1_{(u,v)}.$$

Let $E'$ be the matching of $G_{u,v}$ such that each node in $V'$ is matched to a node in $\{v_2^1, \ldots, v_2^B\} \subseteq V_2$. Since $E_{uv}$ is the minimum perfect matching of $G_{uv}$

$$W(E_{uv}) \leq W(E'). \quad (9)$$

The above inequalities yields

$$f(V'_{u^*,v^*}, u^*, v^*) \leq W(E') + \lambda \cdot B \cdot (N_1 - B) d^1_{(u,v)}. \quad (10)$$

Since $W(E') = \lambda \cdot (N_1 - B) \cdot d^1_{(V',u)} + \lambda \cdot B \cdot d^1_{(X_1 \setminus V', v)} + (1 - \lambda) \cdot d^1_{(X_1 \setminus V'_{u^*,v^*}, X_0)}$, the right side of the above inequality is $h(V', u^*, v^*)$. Lemma 2 thus follows. $\square$

*Theorem 1:* $V'_{u^*,v^*}$ is a 3-approximation of Problem 2.

*Proof:* We simply denote $V'_{u^*v^*}$ by $S'$ and let $V^*$ be the optimal solution to Problem 2. Thus

$$g_1(S') = \lambda \cdot d^1_{(S', X_1/S')} + (1 - \lambda) \cdot d^1_{(X_1/S', X_0)}$$
$$= \lambda \cdot \sum_{u \in S'} \sum_{v \in X_1 \setminus S'} d^1_{(u,v)} + (1 - \lambda) \cdot d^1_{(X_1/S', X_0)}$$
$$\leq \lambda \cdot \sum_{u \in S'} \sum_{v \in X_1 \setminus S'} (d^1_{(u,u^*)} + d^1_{(u^*,v^*)} + d^1_{(v^*,v)})$$
$$+ (1 - \lambda) \cdot d^1_{(X_1/S', X_0)}$$
$$= \lambda \cdot (N_1 - B) \cdot d^1_{(S',u^*)} + \lambda \cdot B \cdot (N_1 - B) \cdot d^1_{(u^*,v^*)}$$
$$+ \lambda \cdot B \cdot d^1_{(v^*, X_1/S')} + (1 - \lambda) \cdot d^1_{(X_1/S', X_0)}$$
$$= h(S', u^*, v^*).$$

By Lemma 2, for every $v \in V^*$ and $u \in X_1 \setminus V^*$

$$f(S', u^*, v^*) \leq f(V^*, u, v)$$
$$= \lambda \cdot (N_1 - B) \cdot d^1_{(V^*, u)} + \lambda \cdot B \cdot d^1_{(v, X_1 \setminus V^*)}$$
$$+ (1 - \lambda) \cdot d^1_{(X_1 \setminus V^*, X_0)} + \lambda \cdot B$$
$$\cdot (N_1 - B) \cdot d^1_{(u,v)}.$$

Adding up the above inequality for all $u \in V^*$ and $v \in X_1 \setminus V^*$, the left side of the summation is

$$h(S', u^*, v^*) \cdot B \cdot (N_1 - B)$$

and the right side is

$$3 \cdot \lambda \cdot B \cdot (N_1 - B) \cdot d^1_{(V^*, X_1 \setminus V^*)} + (1 - \lambda) \cdot B$$
$$\cdot (N_1 - B) \cdot d^1_{(X_1 \setminus V^*, X_0)}$$
$$\leq 3 \cdot B \cdot (N_1 - B) \cdot g_1(V^*).$$

Thus

$$g_1(S') \leq h(S', u^*, v^*) \leq 3 \cdot g_1(V^*). \quad (11)$$

$\square$

As shown in Theorem 2, MBED is a polynomial algorithm with respect to $N$.

*Theorem 2:* MBED algorithm runs in $O(N^2)$.

*Proof:* For each $w \in V$, the first influence distance from $w$ to other nodes can be derived in $O(N^2)$ by the classic shortest path algorithm.[2] Note that we only need to calculate the $d_{(w,v)}$, where $w \in V_1$. Therefore, the calculation of the influence distance totally requires $(N_1 \cdot N^2)$. Given the values of $d_{(w,v)}$, Algorithm 1 takes $O(N_1(B + (N_1 - B)(N - N_1)))$ where calculating $d^1_{(w, X_0)}$ costs $O(N - N_1)$. Furthermore, for each $G_{uv}$, the bipartite minimum perfect matching problem can be solved in $O(N_1^3)$ due to Munkres [20]. Putting these together, Algorithm 2 totally takes $O(N^2 + N_1^2 \cdot (N_1(B + (N_1 - B)(N - N_1)) + N_1^3))$, which is $O(N^2)$ with respect to $N$. $\square$

In fact, MBED is sufficiently fast because $N_1$ is very small compared with $N$. We will later discuss the selection of $\lambda$.

### B. k-IDBSD Problem for $k > 1$

In this section, we show a flow-based heuristic for $k > 1$. We start by transforming Problem 1 to the size-constrained minimum cut problem. In particular, given an IC network $G$ with a target activation state, let us consider a weighted complete graph, denoted by $G^*$, where $V(G^*) = X_1$ and the weight function is defined as follows:

$$w(u, v) = \lambda \cdot d^k_{(u,v)} + (1 - \lambda) \cdot \frac{d^k_{(v, X_0)}}{B}. \quad (12)$$

Similarly, for a node $u \in X_1$ and two subsets $V_1, V_2 \subseteq X_1$, we define that $w(u, V_1) = \sum_{v \in V_1} w(u, v)$ and $w(V_1, V_2) = \sum_{u \in V_1} \sum_{v \in V_2} w(u, v)$. For an edge set $E'$, $W(E') = \sum_{e \in E'} w(e)$. For any two sets $V_1$ and $V_2$ such that $V_1 \cap V_2 = \emptyset$ and $V_1 \cup V_2 = X_1$, we denote by cut$(V_1, V_2)$ the set of cut

---

edges from $V_1$ to $V_2$. Given an effector set $S$ of size $B$, it follows that:

$$W(\text{cut}(S, V(G^*)/S))$$
$$= \sum_{u \in S} \sum_{v \in V(G^*) \setminus S} \lambda \cdot d_{(u,v)}^k + \frac{(1-\lambda) \cdot d_{(v,X_0)}^k}{B}$$
$$= g_k(S). \qquad (13)$$

Therefore, to solve the $k$-IDBSD problem, it is equivalent to finding a subset $S$ of $V(G^*)$ with $B$ nodes such that $W(\text{cut}(S, V(G^*)/S))$ is minimized. Note that if the size restriction on $S$ is removed, it becomes the classic minimum cut problem, which can be efficiently solved by the min-cut max-flow theory. However, if we require that $|S| = B$, it becomes NP-hard. In the following, we provide an efficient algorithm that can produce a node set $S$ with a small $W(\text{cut}(S, V(G^*)/S))$. Our algorithm proceeds in two steps.

*Step 1:* We first obtain the minimum cut of $G^*$,[3] denoting the corresponding partition of $V(G^*)$ by $(S_1, S_2)$. Because $(S_1, S_2)$ may not satisfy the size restriction, we next move the nodes from one set to the other until $|S_1| = B$. In particular, if $|S_1| < B$, we keep moving the node $u$ from $S_2$ to $S_1$ such that $W(\text{cut}(S_1 \cup \{u\}, S_2 \setminus \{u\}))$ is the minimal among all the $u$ in $S_2$, until $|S_1| = B$. We do the same if $|S_1| > B$.[4] By doing this, we now have a new partition $(S_1', S_2')$, where $|S_1'| = B$.

*Step 2:* The second step is to exchange the nodes between $S_1'$ and $S_2'$ such that the objective value is further reduced. In order to show the details, we introduce the following notations. For a pair of nodes $(u_1, u_2)$, where $u \in S_1'$ and $v \in S_2'$, let

$$S_1'(u_1, u_2) = S_1'/\{u_1\} \cup \{u_2\}$$

and

$$S_2'(u_2, u_1) = S_2'/\{u_2\} \cup \{u_1\}.$$

Define the gain by exchanging $u_1$ and $u_2$ by

$$g(S_1', S_2', u_1, u_2) = W(E(S_1', S_2'))$$
$$- W(E(S_1'(u_1, u_2), S_2'(u_2, u_1))).$$

Note that $g(S_1', S_2', u_1, u_2)$ can be negative, which means the current partition cannot be improved by exchanging a single pair of nodes. Thus, a naive approach is to successively exchange a pair of nodes and update the partition until the gain is nonpositive for any pair of nodes. In the following, we show a better searching strategy. We first select the pair of nodes $(u_1^1, u_2^1)$ with a gain $g_1$ such that $g_1$ is maximized among all the choices. Then we repeat this process for $S_1'(u_1, u_2)$ and $S_2'(u_2, u_1)$ where the nodes are selected from $S_1'(u_1, u_2) \setminus \{u_2\}$ and $S_2'(u_2, u_1) \setminus \{u_1\}$. The nodes that have been selected in the past rounds will never be considered again. Thus, the process terminates after $k' = \min(|S_1'|, |S_2'|)$ steps and we will obtain a series of pairs of nodes with the gains $g_1, \ldots, g_{k'}$. Clearly, if we exchange the first $i$ pairs of nodes, the total gain is $\sum_{j=1}^{i} g_j$. Let $k^*$ be the index such that $\sum_{j=1}^{k^*} g_j$ is maximized. If $\sum_{j=1}^{k^*} g_j$ is larger than zero, we exchange the first $k^*$ pairs

---

[3]We can use either the classic Ford–Fulkerson algorithm or the randomized algorithm by Karger and Stein [21].

[4]That is, we move the node from $S_1$ to $S_2$.

---

**Algorithm 3** Flow-Based Effector Detection

1: **Input:** $G = (V, E)$, $X_1$ and $B$.
2: **Output:** A partition of $X_1$.
3: $N_1 := |X_1|$;
4: let $G^*$ be a complete graph with $V(G^*) = X_1$;
5: **for** each pair $u, v \in G^*$ **do**
6: $\qquad w(u, v) = \lambda \cdot d_{(u,v)}^k + (1 - \lambda) \cdot \dfrac{d_{(u,X_0)}^k}{(N_1 - B)}$
7: let $S_1$ and $S_2$ be the partition corresponding to the minimum cut of $G^*$;
8: **if** $|S_1| > B$ **then**
9: $\quad$ **while** $|S_1| > B$ **do**
10: $\qquad u := \arg \min_{u \in S_1} W(\text{cut}(S_1 \setminus \{u\}, S_2 \cup \{u\}))$;
11: $\qquad S_1 := S_1 \setminus \{u\}$;
12: $\qquad S_2 := S_2 \cup \{u\}$;
13: **else**
14: $\quad$ **while** $|S_1| < B$ **do**
15: $\qquad u := \arg \min_{u \in S_2} W(\text{cut}(S_1 \cup \{u\}, S_2 \setminus \{u\}))$;
16: $\qquad S_1 := S_1 \cup \{u\}$;
17: $\qquad S_2 := S_2 \setminus \{u\}$;
18: **while** true **do**
19: $\quad gain(0) := 0$;
20: $\quad gainmax := 0$, $maxindex = 0$;
21: $\quad$ **for** $i = 1 : \min(B, N_1 - B)$ **do**
22: $\qquad (u, v) = \arg \min_{u \in S_1, v \in S_2} W(\text{cut}(S_1 \cup \{u\}, S_2 \setminus \{u\}))$;
23: $\qquad l_{S_1}(i) := u$, $l_{S_2}(i) := v$;
24: $\qquad gain(i) := gain(i - 1) + W(\text{cut}(S_1, S_2)) - W(\text{cut}(S_1 \cup \{u\}, S_2 \setminus \{u\}))$;
25: $\qquad$ **if** $gain(i) > gainmax$ **then**
26: $\qquad\qquad gainmax = gain(i)$, $maxindex = i$;
27: $\quad$ **if** $maxindex > 0$ **then**
28: $\qquad$ **for** $i = 1 : maxindex$ **do**
29: $\qquad\qquad S_1 := S_1 \setminus \{l_{S_1}(i)\} \cup \{l_{S_2}(i)\}$;
30: $\qquad\qquad S_2 := S_2 \setminus \{l_{S_2}(i)\} \cup \{l_{S_1}(i)\}$;
31: $\quad$ **else**
32: $\qquad$ break;
$\quad$ **return** $S_1$

---

of nodes and repeat the whole process by treating the resulting partition as the initial $(S_1', S_2')$, until $\sum_{j=1}^{k^*} g_j$ is less or equal to zero. The partition produced by this step actually reaches the local optimal. The whole process is formally shown in Algorithm 3 named the flow-based effector detection (FBED) algorithm.

## V. Maximum Likelihood Estimation

In this section, we show how to detect the effectors by the MLE approach. That is, searching the node subset that is able to maximize the likelihood of the observed activation state. Let $\Pr(\mathbf{A}^*|S)$ denote the probability of observing $\mathbf{A}^*$ for a given seed set $S$. Therefore, our problem is to find an $S^*$ such that

$$S^* = \arg \max_{S \subseteq X_1, |S| = B} \Pr(\mathbf{A}^*|S). \qquad (14)$$

It is not surprising that given $\mathbf{A}^*$ and $S$, finding such an $S^*$ is an NP-hard problem in general and even calculating the value of $\Pr(\mathbf{A}^*|S)$ is #P-hard. In the following, we first show that

---

**Algorithm 4** Hierarchical Partition

---
1: **Input**: A graph $X_1$
2: **Output**: A partition of $X_1$.
3: $F_0 := \{u | u$ has no parent in $X_1\}$;
4: $V' := X_1 \setminus F_0$;
5: $i := 1$;
6: **while** $V' \neq \emptyset$ **do**
7: $\quad F_i := \{u \in V' | \mathrm{Par}(u) \subseteq \cup_{j=0}^{i-1} F_j\}$;
8: $\quad V' := V' \setminus F_i$;
9: $\quad i := i + 1$;
$\quad$ **return** $F_0, \ldots, F_{i-1}$

---

the MLE can be derived on DAGs in polynomial time, and then discuss the case for general graph.

### A. Directed Acyclic Graph

We start by discussing the case when the social network forms a DAG and $X_1$ is connected. Given an activation state $\mathbf{A}^*$ and an effector set $S$, we will give a closed form of $\mathrm{Pr}(\mathbf{A}^*|S)$ in the following. Since $X_1$ is connected, it must also form a DAG. For a node set $V'$, without causing any confusion, we also refer to $V'$ (also $\overline{V'}$) as the event that the nodes in $V'$ are active (also inactive). Then

$$\mathrm{Pr}(\mathbf{A}^*|S) = \mathrm{Pr}(X_1, \overline{X_0}|S) = \mathrm{Pr}(\overline{X_0}|X_1, S) \cdot \mathrm{Pr}(X_1|S) \quad (15)$$

where $\mathrm{Pr}(\overline{X_0}|X_1, S)$ is the probability that the nodes in $X_0$ are not active on the condition that the nodes in $X_1$ are active for the given seed set $S$. It is easy to verify that

$$\mathrm{Pr}(\overline{X_0}|X_1, S) = \prod_{e \in X_1^{\mathrm{out}}} (1 - \mathrm{Pr}[e]) \quad (16)$$

where $X_1^{\mathrm{out}} = \{(u, v) | u \in X_1, v \in X_0\}$. In order words, given that all the nodes in $X_1$ are activated, nodes in $X_0$ are not activated if and only if all the edges in $X_1^{\mathrm{out}}$ are not live. Because $\mathrm{Pr}(\overline{X_0}|X_1, S)$ is independent of $S$ according to (16), we hence only need to consider the value of $\mathrm{Pr}(X_1|S)$ for maximizing $\mathrm{Pr}(\mathbf{A}^*|S)$. For a node $u$ in $X_1$, we denote the set of the parents of $u$ in $X_1$ by $\mathrm{Par}(u)$. Now let us consider a collection of subsets of $X_1$ returned by Algorithm 4. Let $F_0, \ldots, F_m$ be the output of Algorithm 4. We will in the following show that $F_0, \ldots, F_m$ actually forms a partition of $X_1$. Clearly, $F_i \cap F_j = \emptyset$ for $i \neq j$, and thus, it is sufficient to prove the following lemma, which implies $\cup_{i=0}^m F_i = X_1$.

*Lemma 3:* For each loop from line 6 to 9 in Algorithm 4, if $V' \neq \emptyset$ then $F_i \neq \emptyset$.

*Proof:* For contradiction, suppose $V' \neq \emptyset$ but $F_i = \emptyset$. That is, each node in $V'$ has at least one parent in $V'$. We arbitrarily choose a node in $V'$, namely, $u_0$. Since $u_0$ has at least one parent in $V'$, we arbitrarily select one of them, namely, $u_1$. Since $u_1 \in V'$ and $u_1$ has at least one parent in $V'$, we again select one of its parents in $V'$, namely, $u_2$. By repeating this process, we obtain a sequence of nodes $\{u_i\}$. Since the nodes in $X_1$ are finite, there must be two nodes in $\{u_i\}$ that are the same, which implies that there is a directed cycle in $X_1$. Thus, the proof of Lemma 3 has been proved. $\square$

Since $\{F_0, \ldots, F_m\}$ forms a partition of $X_1$, $\mathrm{Pr}(X_1|S)$ can be decomposed as follows:

$$\begin{aligned}
\mathrm{Pr}(X_1|S) &= \mathrm{Pr}(F_0, \ldots, F_m|S) \\
&= \mathrm{Pr}(F_m|F_0, \ldots, F_{m-1}, S) \cdot \mathrm{Pr}(F_0, \ldots, F_{m-1}|S) \\
&= \prod_{i=0}^m \mathrm{Pr}(F_i|F_0, \ldots, F_{i-1}, S). \quad (17)
\end{aligned}$$

For the nodes in $F_i$, their parents are in $\cup_{j=0}^{i-1} F_j$, and therefore, given the events $F_0, \ldots, F_{i-1}$ and $S$, the nodes in $F_i$ are activated independently from each other. Thus

$$\begin{aligned}
\mathrm{Pr}(F_i|F_0, \ldots, F_{i-1}, S) &= \prod_{u \in F_i} \mathrm{Pr}(u|F_0, \ldots, F_{i-1}, S) \\
&= \prod_{u \in F_i} \mathrm{Pr}(u|\mathrm{Par}(u), S).
\end{aligned}$$

For a node $u \in F_i$

$$\mathrm{Pr}(u|\mathrm{Par}(u), S) = \begin{cases} 1 & \text{if } u \in S \\ 1 - \prod_{v \in \mathrm{Par}(u)} (1 - \mathrm{Pr}[(v, u)]) & \text{if } u \notin S. \end{cases}$$

It follows that:

$$\mathrm{Pr}(F_i|F_0, \ldots, F_{i-1}, S) = \prod_{u \in F_{m-i} \setminus S} \left( 1 - \prod_{v \in \mathrm{Par}(u)} (1 - \mathrm{Pr}[(v, u)]) \right)$$

and by (17)

$$\mathrm{Pr}(X_1|S) = \prod_{u \in X_1 \setminus S} \left( 1 - \prod_{v \in \mathrm{Par}(u)} (1 - \mathrm{Pr}[(v, u)]) \right).$$

Now it is clear that to maximize $\mathrm{Pr}(X_1|S)$ is to find an $S$ that minimizes $\prod_{u \in S}(1 - \prod_{v \in \mathrm{Par}(u)}(1 - \mathrm{Pr}[(v, u)]))$ and it is sufficient to select the nodes that have the smallest value of $1 - \prod_{v \in \mathrm{Par}(u)}(1 - \mathrm{Pr}[(v, u)])$. Note that $1 - \prod_{v \in \mathrm{Par}(u)}(1 - \mathrm{Pr}[(v, u)]) = 0$ if $u \in F_0$. Therefore, $F_0$ must be a subset of the optimal solution. This is intuitive because the nodes in $F_0$ have no parent and thus can only be activated as seed nodes.

The above analysis can be easily generalized to the case when $X_1$ is not connected and consists of isolated DAGs.

### B. General Graph

Generally, the input graph may not be a DAG. For an arbitrary graph, we can first extract a DAG for each connected part of $X_1$ and then apply the above approach to find the maximum likelihood effector set. To avoid introducing new notations, we assume that $X_1$ is connected.

A DAG contained in $X_1$ can be obtained by removing a certain set of edges in $X_1$. It is preferred to extract the DAG that preserves the influence information as much as possible. To this end, we employ the maximum entropy principle. *Entropy* is widely used as a metric of the volume of information contained in a probabilistic system [22]. For an IC model, the entropy $H(D)$ of a DAG $D = (X_1, E)$ is defined as follows:

$$H(D) = -\sum_{e \in E} \mathrm{Pr}[e] \cdot \ln \mathrm{Pr}[e]. \quad (18)$$

---

**Algorithm 5** PBDE

1: **Input:** $G = (V, E)$, $X_1$.
2: **Output:** A edge subset of $E$.
3: Assign an arbitrary total order to $X_1$ and denote the index
    of node $u$ by $I(u)$;
4: $E_L := \emptyset$; $E_R := \emptyset$; $w_L := 0$; $w_R := 0$;
5: **for** each $(u, v)$ in the subgraph induced by $X_1$ **do**
6:     **if** $I(u) > I(v)$ **then**
7:         $E_L := E_L \cup \{(u, v)\}$;
8:         $w_L := w_L - \Pr[(u, v)] \cdot \ln \Pr[(u, v)]$;
9:     **else**
10:        $E_R := E_R \cup \{(u, v)\}$;
11:        $w_R := w_R - \Pr[(u, v)] \cdot \ln \Pr[(u, v)]$;
12: Suppose the entropy of $E_L$ is larger than that of $E_R$;
13: **while** $E_R \neq \emptyset$ **do**
14:     Let $(u, v)$ be the edge in $E_R$ with the largest entropy;
15:     **if** $E_L \cup (u, v)$ is acyclic **then**
16:        $E_L := E_L \cup \{(u, v)\}$;
17:     $E_R := E_R \setminus \{(u, v)\}$;
        **return** the subgraph corresponding to $E_L$;

---

Thus, we aim to remove a set of edges from $X_1$ such that the remaining subgraph forms a DAG that has the maximum entropy according to (18). We call this subproblem the DAG extracting problem. Unfortunately, the DAG extracting problem is NP-hard. Actually, by assigning each edge $(u, v)$ the same propagation probability, it is equivalent to the minimum feedback arc set problem that is to find a set of edges with minimum size that, when removed, leaves a DAG. In this paper, we employ the following approximation algorithm for the DAG extracting problem.

*1) Permutation-Based DAG Extraction [23]:* We first assign an arbitrarily total order (i.e., decide a permutation) of the nodes in $X_1$. Then construct two subgraphs $L = (X_1, E_L)$ and $R = (X_1, E_R)$, where $E_L = \{(u, v) | u > v\}$ and $E_R\{(u, v) | u < v\}$. Without loss of generality, assume $L$ has larger entropy. Now we move the edge from $R$ to $L$ such that $L$ remains acyclic until no such edge exists and output $L$. When moving edge from $R$ to $L$, we first check the edge that has the largest value of $-\Pr[e] \cdot \ln \Pr[e]$. We denote this algorithm by the permutation-based DAG extraction (PBDE) algorithm, which is described in Algorithm 5.

*Theorem 3:* The PBDE algorithm finds a DAG in $X_1$ that has an entropy at least half of that of the optimal DAG.

*Proof:* First, there is no directed cycle in $L$ because $u > v$ for each edge $(u, v)$ in $L$. Second, the entropy of $L$ is at least half of the optimal solution because

$$H(L) \geq (H(L) + H(R))/2 \geq H(D_{\text{opt}})/2$$

where $D_{\text{opt}}$ is the optimal solution. $\square$

As mentioned earlier, the nodes that have no parent in $X_1$ must be the effectors, for otherwise $\Pr(X_1|S) = 0$. Thus, the last step of PBDE algorithm is important as it is can guarantee such nodes to be included in the extracted DAG. Other sophisticated algorithms for the DAG extracting problem can be obtained from [23] and [24] regarding the minimum feedback arc set problem. We denote the above approach by the maximum-likelihood-based effector detection (MLBED) algorithm. Note that the MLBED approach also can be used to solve the influence source detection problem.

## VI. EXPERIMENTS

In this section, we experimentally evaluate the proposed methods for effector detection.

### A. Experimental Setup

We use the Facebook network [19], which has been widely adopted for studying social networks [25]–[27]. The selected Facebook network includes 4039 users and 88234 undirected edges. We replace each edge with two directed edges to make the graph directed. We consider two popular settings of propagation probability.

1) *Uniform Setting:* $\Pr[(u, v)] = 0.01$ for each edge. Such a setting represents the case when we lack the knowledge of the given social network.[5]
2) *Weighted Cascade Setting:* $\Pr[(u, v)] = 1/\deg(v)$, where $\deg(v)$ is the number of nodes connected to $v$. As mentioned in [10], the high degree nodes are too influential under the uniform setting, and the weighed cascade setting balances the degree and the propagation probability.

For each activation state, we obtain the effectors according to the algorithms and calculate $f_1(S)$ via Monte Carlo simulations.[6] By Hoeffding's inequality, the produced $f_1(S)$ can be as accurate as possible when a sufficient number of simulations have been performed. We herein set the number of iterations as 10 000. The considered effector detection algorithms are as follows.

1) *MBED:* This is the effector detection algorithm designed in Section IV-A. MBED searches good effectors by solving Problem 2 formulated based on the first influence distance.
2) *FBED:* This is the effector detection algorithm designed in Section IV-B. FBED searches good effectors by solving Problem 1 formulated based on the $k$th influence distance. In particular, we set $k = 3$.
3) *MLBED:* This is the effector detection algorithm designed in Section V. MLBED searches the effectors that can maximize the likelihood of the target activation state.
4) *OutDegree:* This is a heuristic algorithm proposed in [2] that selects effectors with the highest out-degree in the influence tree. The details of this algorithm can be found in [2].
5) *Random:* This is a baseline algorithm that randomly selects effectors from the active nodes.

Note that the other methods proposed in [2] are not included in our experiment because their performances are similar to that of OutDegree as shown there.

---

[5]Deriving the edge probabilities is beyond the scope of this paper.
[6]We do not report the result under $f_2(S)$ because the experimental results under $f_1(S)$ and $f_2(S)$ are similar.

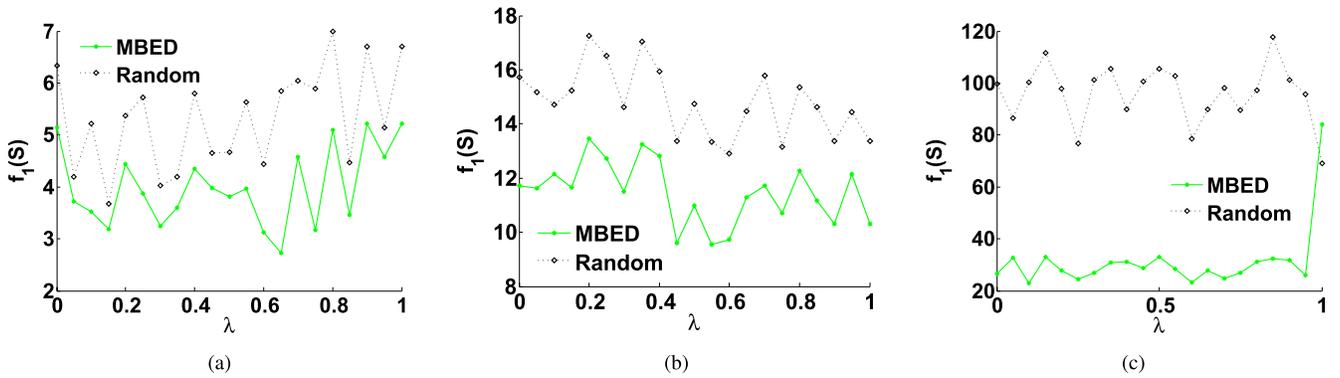

Fig. 3. Performance of MBED under different $\lambda$. The $y$-axis and $x$-axis denote the value of $\lambda$ and the score $f_s(S)$, respectively. Each graph gives two curves showing the performance of MBED and Random, respectively. (a) $N_1 = 5$ under uniform setting. (b) $N_1 = 15$ under weighted cascade setting. (c) $N_1 = 24$ under weighted cascade setting.

## B. Result

In this section, we show the results of the two experiments with different methods for generating activation states.

*1) First Experiment:* In this experiment, given the budget $B$, we first randomly select $B$ nodes as seed nodes and then generate activation states by simulating the diffusion process.

First, we briefly discuss the selection of $\lambda$. According to the formulation, the selection of $\lambda$ intuitively depends on $N_1$ (i.e., the number of active nodes in $\mathbf{A}^*$). However, we observe that as long as $\lambda$ is less than 1, the performance of MBED does not vary much. For a fixed activation state, we apply MBED tuning $\lambda$. Fig. 3 shows several results for different $N_1$. For the first two cases shown in Fig. 3(a) and (b), the performance is slightly better when $\lambda$ is close to 0.5. However, when $N_1$ is further increased, the performances of MBED are in the same level with different $\lambda$ less than 1.

Second, let us discuss the effectiveness of the considered effector detection methods. First, we consider the case of single effector, i.e., $B = 1$. As shown in Table II, in most cases, MBED and FBED have better performance than MLBED and Random. Recall that the small $f_1(S)$ implies the high quality of the effectors. However, the difference of the metric score between the effectors produced by different algorithms is not significant because one seed node cannot result in many active nodes, i.e., $N_1$ is small. Next, we set $\lambda$ to be 0.5 and collect the results for larger $N_1$. The results under the uniform setting and weighted cascades are shown in Fig. 4(a) and (b), respectively. The details of the result in Fig. 4(a) are shown in Table III in Appendix. From Fig. 4, the following can be observed.

1) MBED provides the best performance in most cases.
2) FBED and MLBED perform slightly worse than MBED does.
3) MBED, FBED, and MLBED are superior to OutDegree by a significant gap.

Now, let us discuss the intuitions behind those results. The MBED is highly effective because it is able to minimize $g_1(S)$ with a theoretical guarantee and $g_1(S)$ is linearly correlated to $f_1(S)$ as shown in Fig. 2. Although FBED does not have an approximation ratio for minimizing $g_k(S)$, it can utilize the $k$th influence distance that captures more information of the influence diffusion between the nodes. Because MLBED finds

## TABLE II
### FIRST EXPERIMENT WITH $B = 1$ UNDER UNIFORM SETTING

| $\lambda$ | MBED | FBED | MLBED | OutDegree | Random |
|---|---|---|---|---|---|
| 0.05 | **1.969** | 2.009 | 2.164 | 2.147 | 2.287 |
| 0.1 | **2.767** | 3.062 | 3.570 | 3.143 | 3.967 |
| 0.15 | 2.471 | 2.724 | 3.466 | **2.231** | 3.826 |
| 0.2 | 2.957 | 3.055 | 3.424 | **2.120** | 3.701 |
| 0.25 | **2.699** | 2.896 | 3.017 | 2.721 | 3.264 |
| 0.3 | **3.377** | 3.447 | 3.744 | 3.761 | 5.102 |
| 0.35 | 1.657 | 1.767 | 1.882 | **1.321** | 2.894 |
| 0.4 | 2.598 | **2.478** | 3.063 | 2.673 | 3.347 |
| 0.45 | 2.077 | **2.008** | 2.016 | 2.103 | 2.332 |
| 0.5 | 3.592 | **3.467** | 4.541 | 4.230 | 4.706 |
| 0.55 | 3.705 | 3.650 | **3.149** | 3.253 | 4.058 |
| 0.6 | 2.701 | 2.291 | 2.034 | **2.023** | 2.803 |
| 0.65 | 2.723 | 2.939 | 3.437 | **2.721** | 3.632 |
| 0.70 | 2.438 | **2.207** | 2.879 | 2.564 | 3.526 |
| 0.75 | **1.776** | 1.778 | 1.888 | 1.793 | 2.063 |
| 0.80 | 2.489 | **2.442** | 3.085 | 2.846 | 3.564 |
| 0.85 | 1.847 | **1.803** | 2.000 | 1.963 | 2.241 |
| 0.90 | **1.911** | 2.103 | 1.924 | 2.023 | 2.944 |
| 0.95 | 4.333 | **4.179** | 4.290 | 4.653 | 4.868 |

the seed nodes with the maximum likelihood, it is suitable for the activation states that are generated by the seed nodes. Therefore, MLBED also performs well in this experiment. According to Table II and Fig. 4(a), the OutDegree algorithm is comparable to other algorithms only if $|X_1|$ is small. It is worth noting that none of the considered algorithms can strictly dominate others, implying that a combination of the proposed methods could be a good choice in practice.

Another important observation is that the quality of the effector heavily depends on the pattern of the activation state, and one can imagine that there exists some activation state such that no effector set can have high quality. Consider the illustration graph shown in Fig. 5. Suppose the number of effector is one. Such an activation state is somehow elusive for finding effector because each active node has a low probability to activate its active neighbors but a high probability to activate its inactive neighbor. Therefore, it is not possible to identify good effectors solely based on the network topology. Such a scenario explains that why OutDegree may have the same degree of effect as that of MBED for a certain activation state.

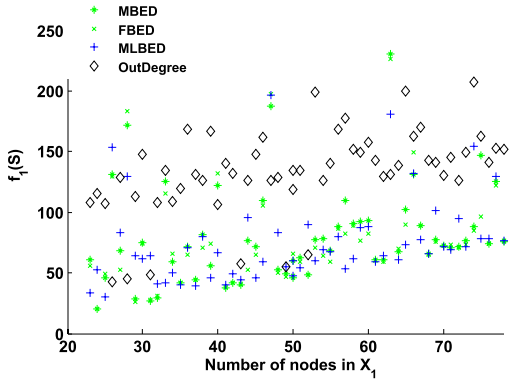

(a)

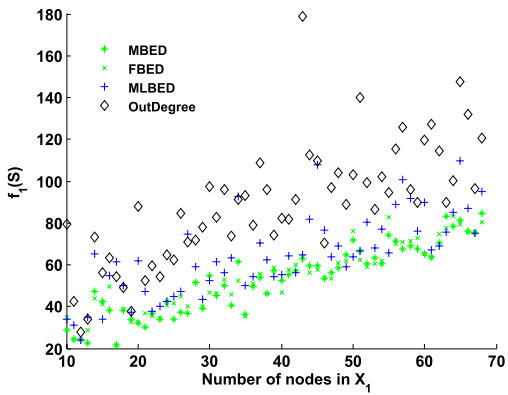

(b)

Fig. 4. Experimental results of the first experiment. The *y*-axis and *x*-axis denote the value of $|X_1|$ and the score $f_1(S)$, respectively. Each graph gives four kinds of markers showing the performance of MBED, FBED, MLBED, and OutDegree, respectively. (a) Uniform setting. (b) Weighted cascade setting.

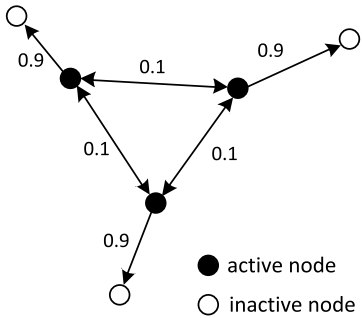

Fig. 5. Elusive activation state.

*2) Second Experiment:* In the second experiment, we first set the value of $|X_1|$ and then select the nodes for $X_1$ from $V$ in random. The budget $B$ is randomly selected from $0.1 \cdot |X_1|$ to $0.2 \cdot |X_1|$. The experimental results are shown in Fig. 6(a) and (b). Not surprisingly, the pattern of the scatter plots is similar to that of the first experiment and the proposed algorithms are still effective. An important observation is that the MLBED algorithm does not perform so well as it does in the first experiment. For the uniform setting, the performances of MLBED and MBED are very close to each other in the first experiment as shown in Fig. 4(a) and Table III in Appendix, while MBED is clearly better than MLBED in the second experiment as shown in Fig. 6(b). Similarly, the difference in the performances between MLBED and MBED in this experiment becomes larger than it is in the first experiment. This is because the MLBED algorithm is designed mainly based on the idea of finding the influence source, but the activation states in this experiment may not be a valid input for the influence source detection problem. Such an observation also implies that the influence source detection methods are feasible but may not perform well for solving the effector detection problem.

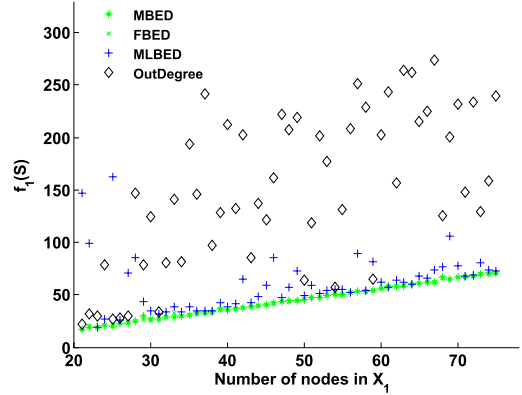

(a)

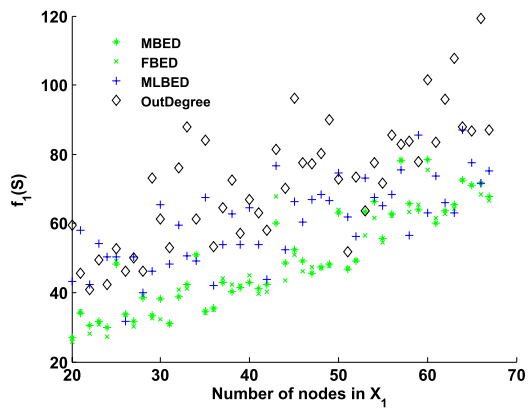

(b)

Fig. 6. Experimental results of the second experiment. (a) Uniform setting. (b) Weighted cascade setting.

## VII. CONCLUSION

In this paper, we have considered the problem of finding effectors in social networks. We first design an effector detection framework based on the idea of influence distance. Specifically, we show how to extract the $k$th influence distance from the IC model and then provide a 3-approximation algorithm for $k = 1$ and a heuristic for $k \geq 1$, respectively. Then we show how the effectors can be detected by the MLE. For DAG graphs, we provide an algorithm that finds the optimal MLE in polynomial time. For general graphs, we extract the DAG subgraph based on the maximum entropy principle and then apply the MLE approach. For the future work, we plan to analyze the effector detection problem for other information-propagation models such as a linear threshold model. As discussed in Section VI, for some activation states, it is not possible to find good effectors. Thus, it is interesting to investigate whether a meaningful effector exists for a given activation state.

TABLE III

TABLE OF FIG. 4(a)

| $B$ | MBED | FBED | MLBED | OutDegree |
|-----|------|------|-------|-----------|
| 50 | **46.32** | 47.54 | 47.85 | 134.44 |
| 51 | 62.68 | 59.31 | **54.16** | 134.53 |
| 52 | **48.43** | 49.12 | 89.96 | 64.78 |
| 53 | 77.54 | 70.85 | **59.88** | 199.47 |
| 54 | 78.38 | **64.05** | 69.36 | 126.47 |
| 55 | 68.24 | **59.00** | 67.14 | 140.21 |
| 56 | 87.83 | 86.33 | **80.28** | 168.81 |
| 57 | 109.89 | 82.28 | **53.71** | 177.66 |
| 58 | 90.76 | 89.01 | **61.53** | 152.14 |
| 59 | 92.15 | **76.46** | 87.00 | 149.81 |
| 60 | 93.36 | **82.84** | 88.07 | 157.98 |
| 61 | 60.62 | 60.18 | **59.55** | 143.21 |
| 62 | **56.79** | 59.34 | 64.04 | 129.30 |
| 63 | 230.67 | 226.27 | **181.02** | 131.62 |
| 64 | 67.46 | 70.41 | **61.29** | 139.11 |
| 65 | 102.13 | 90.29 | **72.90** | 199.98 |
| 66 | **131.22** | 149.62 | 131.74 | 162.81 |
| 67 | 88.95 | 89.35 | **77.79** | 170.32 |
| 68 | 66.25 | **64.72** | 65.89 | 143.20 |
| 69 | 77.82 | **75.63** | 101.42 | 141.27 |
| 70 | 72.66 | 72.48 | **71.61** | 130.33 |
| 71 | 70.88 | 73.65 | **69.56** | 145.53 |
| 72 | 67.70 | **67.53** | 69.13 | 234.21 |
| 73 | **69.23** | 69.44 | 80.60 | 129.45 |
| 74 | **68.43** | 69.88 | 73.97 | 158.98 |
| 75 | **70.84** | 71.10 | 72.26 | 239.90 |

## Appendix
### Detailed Experimental Results

See Table III.

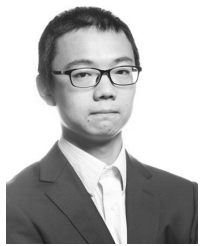

**Guangmo (Amo) Tong** (S'14) received the B.S. degree in mathematics and applied mathematics from the Beijing Institute of Technology, Beijing, China, in 2013, and the Ph.D. degree from the Department of Computer Science, University of Texas, Dallas, TX, USA, under the supervision of Dr. D.-Z. Du and Dr. C. Liu.

He has authored several papers on prestigious conferences and journals. His current research interests include social networks, bigdata analysis, and real-time systems.

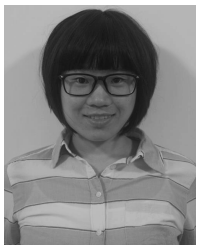

**Sasha Li** received the Ph.D. degree in applied mathematics from the Center for Combinatorics, Nankai University, Tianjin, China, in 2012.

She is currently a Lecturer with the Ningbo Institute of Technology, Zhejiang University, Ningbo, China. Her current research interests include graph theory and its applications, combinatorial optimizations, and algorithms and complexity analysis.

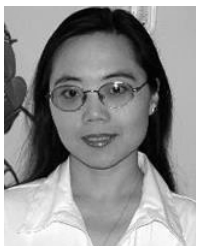

**Weili Wu** (M'00) received the M.S. and Ph.D. degrees from the Department of Computer Science, University of Minnesota, Minneapolis, MN, USA, in 1998 and 2002, respectively.

She is currently a Full Professor with the Department of Computer Science, University of Texas, Dallas, TX, USA. She is involved in the design and analysis of algorithms for optimization problems that occur in wireless networking environments and various database systems. Her current research interests include data communication and data management.

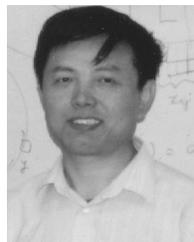

**Ding-Zhu Du** received the M.S. degree from the Chinese Academy of Sciences, Beijing, China, in 1982, and the Ph.D. degree from the University of California, Santa Barbara, CA, USA, in 1985, under the supervision of Prof. R. V. Book.

Before settling at the University of Texas, Dallas, TX, USA, he was a Professor at the Department of Computer Science and Engineering, University of Minnesota, Minneapolis, MN, USA. He was also with the Mathematical Sciences Research Institute, Berkeley, CA, USA, for a year, then with the Department of Mathematics, Massachusetts Institute of Technology, Cambridge, MA, USA, for a year, and later with the Department of Computer Science, Princeton University, Princeton, NJ, USA, for one and a half years. He is the Editor-in-Chief of the *Journal of Combinatorial Optimization* and is also on the editorial boards for several other journals. He has also supervised 40 Ph.D. students.